# Decoding Stimulus Reconstruction-Based Auditory Attention Robustly in Unbalanced EEG Datasets

Yuanming Zhang, Yayun Liang, Zhibin Lin, Jing Lu

*Abstract*— **In the past decade, numerous studies have applied deep neural networks (DNNs) to decode auditory attention (AAD) from Electroencephalogram (EEG) signals via stimulus reconstruction. However, the influence of dataset balance on the decoding performance of stimulus reconstruction-based AAD remains unexplored. In this study, three publicly available EEG-AAD datasets — KUL, DTU, and NJU cEEGrid — are used to construct both balanced and unbalanced experimental conditions. We hypothesize and demonstrate that stimulus reconstruction-based DNN decoders tend to produce overestimated decoding performance on unbalanced datasets. To address this issue, we propose a leave-one-paired-envelope-out (LOPEO) cross-validation protocol. Experimental results confirm that LOPEO effectively prevents inflated decoding accuracy on unbalanced datasets. While balanced datasets are generally preferred in experimental design, LOPEO provides a principled evaluation framework for unbalanced datasets that have already been published, filling an important gap in the field.**

## I. INTRODUCTION

Humans possess an ability to selectively focus on a particular voice while filtering out background noise, a phenomenon known as the cocktail party effect [1]. Selective auditory attention manifests as distinct neural activities, which can be non-invasively captured by Electroencephalogram (EEG) devices [2], [3], [4].

This work is supported by XXX.

Yuanming Zhang and Jing Lu are with Key Lab of Modern Acoustics, Nanjing University, Nanjing 210093, China, and also with the NJU-Horizon Intelligent Audio Lab, Horizon Robotics, Beijing 100094, China (e-mail: yuanming.zhang@smail.nju.edu.cn; lujing@nju.edu.cn)

Yayun Liang and Zhibin Lin is with Key Lab of Modern Acoustics, Nanjing University, Nanjing 210093, China (e-mail: yayun.liang@smail.nju.edu.cn; zblin@nju.edu.cn).

While DNN-based stimulus reconstruction AAD has shown great promise in recent years [5], [6], [7], [8], [9], [10], several studies have highlighted potential overestimation of decoding performance when DNN models are trained and evaluated inappropriately [11], [12], [13], [14]. Leave-one-trial-out (LOTO) has since become the commonly used cross-validation strategy to mitigate this issue.

Although numerous EEG-AAD datasets have been developed to train and evaluate DNN models [11], [15], [16], [17], [18], [19], [20], the impact of dataset balance-specifically, whether each audio stimulus serves equally often as the attended and unattended stream-on decoding performance has not been systematically examined.

Given the limited size of available EEG-AAD datasets and the strong fitting capacity of DNN models, we hypothesize that stimulus reconstruction-based AAD decoders may produce overestimated accuracy when trained and tested on unbalanced datasets. We formalize this hypothesis via a balance index (BI), demonstrate overestimation on three public datasets using the VLAAI model [21], and propose leave-one-paired-envelope-out (LOPEO) cross-validation to mitigate the effect. It should be noted that the scope of this work is restricted to the stimulus reconstruction paradigm; whether similar overestimation occurs in classification-based AAD paradigms remains an open question.

The main contributions are: (1) a formal definition of dataset balance for EEG-AAD via the BI; (2) empirical validation that stimulus reconstruction-based DNN decoders overestimate accuracy on unbalanced datasets; (3) the proposal of the LOPEO cross-validation strategy to mitigate overestimation in unbalanced datasets.

The remainder of this paper is organized as follows. Section II describes dataset construction. Section III presents the LOPEO strategy and training details. Section IV presents results and analysis. Section V concludes with recommendations.

## II. DATASETS AND DATA PREPARATIONS

### A. Definition of Balanced and Unbalanced Datasets

Dataset balance is quantified by the frequency with which each audio stimulus appears as attended speech versus the unattended speech. A dataset is considered balanced when all audio stimuli appear equally often in both roles. Conversely, an unbalanced dataset contains stimuli that appear disproportionately in one role.

*Table 1 Dataset summary. The number of speakers represents simultaneous competing speakers presented to the participants.*

| Dataset | # trials | # subjects | # speakers | Balance Index | Chance Level |
|---|---|---|---|---|---|
| KUL | 20 | 16 | 2 | 0.600 | 0.5 |
| DTU | 60 | 18 | 2 | 0.056 | 0.5 |
| NJU cEEGrid | 63 | 98 | 3 | 0.185 | 1/3 |

We define the balance index (BI):

$$BI = \frac{1}{N_{audio}} \sum_{j=1}^{N_{audio}} \frac{|n_j^{att} - n_j^{unatt}|}{n_j^{att} + n_j^{unatt}} \quad (1),$$

where $N_{audio}$ is the total number of unique audio streams, $n_j^{att}$ and $n_j^{unatt}$ are the counts of the $j$-th stimulus appearing as attended and unattended, respectively. Since $|n_j^{att} - n_j^{unatt}| \le n_j^{att} + n_j^{unatt}$, each term lies in $[0,1]$, and therefore $0 \le BI \le 1$. $BI = 0$ indicates a perfectly balanced dataset; $BI = 1$ indicates extreme imbalance (each stimulus appears exclusively in one role). For three-speaker trials (NJU cEEGrid), each trial contributes one attended count and one unattended count per stimulus involved, preserving the formula's validity.

### B. Dataset Summary

EEG preprocessing for all datasets follows the respective original publications [17], [18], [19]. Table 1 summarizes the three datasets used in this paper.

### C. Constructing Balanced and Unbalanced Datasets

KUL: A balanced dataset (BI = 0) is constructed by selecting the first 8 trials per subject. An unbalanced dataset (BI = 1) includes trial 1 to trial 4, retaining only trials where one speaker is always attended.

DTU: Half of the audio streams are manually selected; only trials with attended audio in which the attended audio belongs to the selected set are retained. The balance index is 1. The number of cross-validation folds is doubled so that all audio streams are represented equally across folds.

NJU cEEGrid: Trials sharing the same attended speaker are retained. Because different speakers narrate different news segments, excluding certain attended speakers yields a dataset with no overlap between attended and unattended audio streams, giving BI = 1. The number of cross-validation folds is extended so that all attended speakers are represented equally.

## III. Cross-validation And Model Training

### A. Task Description

The goal of stimulus reconstruction-based AAD is to predict the attended speech envelope from synchronized EEG signal. Given DNN parameters $\mathbf{\Theta}$ and model $f$, the task can be written as

$$\hat{\mathbf{y}} = f(\mathbf{x} \mid \mathbf{\Theta}) \quad (2),$$

where $\mathbf{x}$ is the input EEG segment, and $\hat{\mathbf{y}}$ is the predicted envelope. The decoding accuracy is defined as

$$\mathrm{Acc} = \frac{1}{|\mathcal{T}_{test}|} \sum_{t \in \mathcal{T}_{test}} \mathbf{1}[\rho(\hat{\mathbf{y}}_t, \mathbf{y}_t^{att}) > \rho(\hat{\mathbf{y}}_t, \mathbf{y}_t^{unatt})] \quad (3).$$

The optimization objective is to minimize the negative Pearson correlation coefficient (PCC) between the predicted envelope and the attended envelope $\mathbf{y}^{att}$.

$$\mathcal{L}_{PCC} = -\rho(\hat{\mathbf{y}}, \mathbf{y}^{att}) \quad (4),$$

where $\rho(x, y)$ represents the PCC function. Additionally, we employ a revised version of Eq. 3, the contrastive PCC loss, as another training goal:

$$\mathcal{L}_{\Delta PCC} = -\rho(\hat{\mathbf{y}}, \mathbf{y}^{att}) + \frac{1}{N_{speaker} - 1} \sum_{k=1}^{N_{speaker}} \rho(\hat{\mathbf{y}}, \mathbf{y}_k^{unatt}) \quad (5),$$

where $\mathbf{y}_k^{unatt}$ is the speech envelope narrated by the k-th unattended speaker.

### B. Leave-one-trial-out Cross-validation

Leave-one-trial-out (LOTO) cross-validation has been widely used in recent AAD research [22], [23], [24]. In LOTO, the training, validation, and test set contain unique, non-overlapped EEG trials. LOTO was originally proposed to mitigate overestimation caused by EEG long range temporal correlation [11], [13]. However, although LOTO prevents EEG segment overlap between data splits, it does not prevent the same audio stimulus from appearing in both the training and test sets.

### C. Leave-one-paired-envelope-out Cross-validation

LOPEO imposes a strictly stronger constraint: the entire unordered stimulus pair (attended stimulus, unattended stimulus) of each test trial must be absent from the training and validation sets.

Formally, any valid LOPEO partition is also a valid LOTO partition, but not vice versa. This additional constraint prevents the decoder from utilizing stimulus-identity-specific features learned during training. Algorithm 1 demonstrates the pseudo-code of LOPEO strategy.

LOEO for NJU cEEGrid: The NJU cEEGrid dataset contains 3-speaker trials. Since the attended speakers cannot be reliably assigned to a fixed pair (different co-speaker combinations appear across trials), it is much harder to form stable attended–unattended pairs for partitioning. We therefore apply a leave-one-envelope-out (LOEO) variant, in which only the attended stimulus is used as the partitioning criterion. LOEO is a weaker version of LOPEO that controls for attended-stimulus identity but not competing-stimulus identity. This is acknowledged as a limitation; future datasets with fixed speaker pairings would enable full LOPEO application.

### D. Training Procedure

The hyperparameters of the VLAAI network are set identical to its original configuration [21]. SuperHugeAAD[1], a Python package built upon PyTorch and PyTorch Lightning, is used to manage the training and evaluation. The Adam optimizer is used with a learning rate of 0.0005 and weight decay of 0.0005 [25]. Learning rate is reduced with a factor of 0.5, patience of 5 epochs, and cooldown of 5 epochs. An early stopping strategy with a patience of 10 epochs is applied. The maximum number of training epochs is set to 100.

[1] https://github.com/SeanZhang99/SuperHugeAAD

The model is trained for each partition defined by the training, validation, and test sets (the output of Algorithm 1). The reported results are averaged across all folds. Full training scripts and logs will be made available upon acceptance of this paper.

## IV. Results

### A. Comparison of Balanced and Unbalanced Datasets

Table 2 presents the decoding performance of the VLAAI model. By comparing Line 1-3, Line 4,5, and Line 6,7, it is clearly shown that the VLAAI model achieved significantly higher decoding accuracy on more unbalanced datasets (higher BI) under both $\mathcal{L}_{PCC}$ and $\mathcal{L}_{\Delta PCC}$ loss functions.

Statistical significance is assessed using two-sided paired Wilcoxon signed-rank tests (unit of analysis: per fold), with Bonferroni correction. The accuracy difference between BI=0 and BI=1 is significant in KUL ($p<0.01$) and NJU cEEGrid ($p<0.001$), but not significant for DTU ($p = 0.12$ for the PCC loss; $p = 0.08$ for the contrastive loss).

In a dataset with BI = 1, attended envelopes never appear as unattended envelopes and vice versa. Under LOTO, the same attended envelope can appear in both the training and test sets. The DNN therefore learns stimulus-identity-specific features tied to the attended envelope, which inflates $\rho_a$ and thus accuracy, without reflecting any genuine improvement in attention decoding. The contrastive loss amplifies this effect further by explicitly maximizing $\rho_a - \rho_u$.

The DTU dataset has 60 unique audio stimuli (one per trial per subject), meaning that each audio stimulus appears only once per subject. The ratio of unique stimulus pairs to total trials is therefore 1:1. Therefore, even under LOTO there is naturally limited pair-level repetition. This structural property inherently reduces the opportunity for stimulus-identity leakage, explaining the weaker overestimation observed in DTU dataset. Moreover, the stimuli pairings in the DTU dataset are not fixed across subjects. In contrast, KUL has only 8 unique audio streams shared across 20 trials, creating substantial repetition.

It is noted that the current experiments use VLAAI only. The overestimation mechanism likely arises from the memorization of features of attended envelopes, which requires sufficient model capacity[2]. A linear baseline such as ridge regression would not exhibit the same degree of memorization and thus may show weaker or no overestimation. Testing additional architectures in future work also remains an open question.

### B. Evaluating Decoders with Left-out Envelopes

Starting from Line 15 of Table 2, a notable effect on decoding performance is observed when the LOPEO strategy is applied. Specifically, decoding accuracy on KUL dataset with different Balance Index consistently falls within a similar range. The decoder achieves similar decoding performance across the different variations of the DTU dataset when $\mathcal{L}_{PCC}$ is used. These results confirm that LOPEO prevents DNN decoders from exploiting stimulus-identity leakage.

The decoding performance on NJU cEEGrid dataset decreases substantially when the LOEO strategy is applied. Only at Line 28 does decoding accuracy exceed the chance level (1/3). Under LOEO, when the attended stimulus is withheld from training, the model may learn spurious correlations with

*Algorithm 1 Leave-one-paired-envelope-out Cross-Validation*

1 **Input**: EEG trials set $D=\{d_1,d_2,...,d_n\}$, attended stimuli set $S_a=\{s_{a,1},s_{a,2},...,s_{a,n}\}$, unattended stimuli set $S_u=\{s_{u,1},s_{u,2},...,s_{u,n}\}$, number of cross-validation folds $K$. A sorting algorithm $\mathbf{f}(x,y)$ that ensures $\mathbf{f}(s_{a,i},s_{u,i})\equiv\mathbf{f}(s_{u,i},s_{a,i}),\mathbf{f}(s_{a,i},s_{u,i})\neq\mathbf{f}(s_{a,i},s_{u,j})$ $\forall s_{a,i}\in S_a, s_{u,i}\in S_u, s_{u,j}\in S_u \,\&\, s_{u,j}\neq s_{u,i}$.

2 **Output**: Training, validation, and test set partitions of each fold $(D_{train},D_{val},D_{test})$.

3 Initialize <u>unordered</u> stimuli pair set $S_{pair}=\varnothing$.

4 **for each** trial $d_i\in D$ **do**

5 Get the unordered stimuli pair $\mathbf{f}(s_{a,i},s_{u,i})$ of trial $d_i$.

6 $S_{pair}\leftarrow S_{pair}\cup\{\mathbf{f}(s_{a,i},s_{u,i})\}$.

7 **end for**

8 Randomly partition $S_{pair}$ into $K$ folds: $S_1,S_2,...,S_K$.

9 **for each** test fold $S_t, t=1,2,...,K$, **do**

10 **for each** validation fold $S_v, v=1,2,...,K, v\neq t$, **do**

11 Find training fold set $S_r=\{S_i \,\|\, i\neq t \,\&\, i\neq v\}$.

12 Initialize $D_{train},D_{val},D_{test}=\varnothing$

13 **for each** trial $d_i\in D$, **do**

14 Get the unordered stimuli pair $\mathbf{f}(s_{a,i},s_{u,i})$ of trial $d_i$.

15 **if sorted**$(s_{a,i},s_{u,i})\in S_r$ **then**

16 $D_r\leftarrow D_r\cup\{d_i\}$

17 **else if sorted**$(s_{a,i},s_{u,i})\in S_v$ **then**

18 $D_v\leftarrow D_v\cup\{d_i\}$

19 **else if sorted**$(s_{a,i},s_{u,i})\in S_t$ **then**

20 $D_t\leftarrow D_t\cup\{d_i\}$

21 **end if**

22 **end for**

23 **yield** $(D_{train},D_{val},D_{test})_{r,t,v}$

24 **end for**

25 **end for**

[2] Experiment results supplied in additional materials.

*Table 2 Experiment results under leave-one-trial-out cross-validation and leave-one-paired-envelope-out strategy. Acc represents the decoding accuracy. $\rho_a, \rho_u$ is the Pearson correlation coefficient between the reconstructed envelope and the attended/unattended envelopes, respectively. $\Delta\rho = \rho_a - \rho_u$ represents the differences between attended and unattended PCC. Balance Index (BI) represents the balance of a dataset. A dataset with a BI of 0 is well-balanced. A dataset with a BI of 1 is extremely unbalanced. The decoding performance is averaged across all folds.*

| Line # | Cross-validation Strategy | Dataset | Chance Level (Acc) | Balance Index | Loss | Acc | $\rho_a$ | $\rho_u$ | $\Delta\rho$ |
|---|---|---|---|---|---|---|---|---|---|
| 1 | Leave-one-trial-out | KUL | 0.500 | 0 | $\mathcal{L}_{PCC}$ | 0.6493±0.0156 | 0.0984±0.0025 | 0.0499±0.0061 | 0.0485±0.0054 |
| 2 | | KUL | | 0.600 | | 0.6895±0.0065 | 0.1124±0.0018 | 0.0488±0.0033 | 0.0636±0.0022 |
| 3 | | KUL | | 1 | | 0.8319±0.0101 | 0.1335±0.0031 | 0.0153±0.0032 | 0.1182±0.0050 |
| 4 | | DTU | | 0.056 | | 0.6527±0.0206 | 0.1170±0.0065 | 0.0272±0.0032 | 0.0899±0.0080 |
| 5 | | DTU | | 1 | | 0.6823±0.0269 | 0.1233±0.0120 | 0.0168±0.0056 | 0.1066±0.0098 |
| 6 | | NJU cGrid | 0.333 | 0.185 | | 0.3398±0.0086 | 0.0748±0.0008 | 0.0706±0.0009 | 0.0043±0.0016 |
| 7 | | NJU cGrid | | 1 | | 0.5951±0.0378 | 0.1146±0.0111 | 0.0431±0.0074 | 0.0714±0.0124 |
| 8 | | KUL | 0.500 | 0 | $\mathcal{L}_{\Delta PCC}$ | 0.6650±0.0104 | 0.0581±0.0047 | -0.0009±0.005 | 0.0589±0.0037 |
| 9 | | KUL | | 0.600 | | 0.7171±0.0111 | 0.0654±0.0020 | -0.0217±0.0046 | 0.0871±0.0040 |
| 10 | | KUL | | 1 | | 0.8960±0.0097 | 0.0921±0.0035 | -0.068±0.0031 | 0.1601±0.0054 |
| 11 | | DTU | | 0.056 | | 0.5623±0.0217 | 0.0453±0.0101 | 0.0034±0.0055 | 0.0419±0.0122 |
| 12 | | DTU | | 1 | | 0.6393±0.0186 | 0.0636±0.0082 | -0.0191±0.0022 | 0.0827±0.0069 |
| 13 | | NJU cGrid | 0.333 | 0.185 | | 0.3730±0.0044 | 0.0137±0.0025 | 0.0020±0.0024 | 0.0117±0.0020 |
| 14 | | NJU cGrid | | 1 | | 0.6765±0.0220 | 0.0748±0.0158 | -0.0315±0.0165 | 0.1063±0.0058 |
| 15 | Leave-one-paired-envelope-out | KUL | 0.500 | 0 | $\mathcal{L}_{PCC}$ | 0.6493±0.0156 | 0.0984±0.0025 | 0.0499±0.0061 | 0.0485±0.0054 |
| 16 | | KUL | | 0.600 | | 0.6420±0.0143 | 0.0707±0.0036 | 0.0240±0.0029 | 0.0467±0.0045 |
| 17 | | KUL | | 1 | | 0.6467±0.0512 | 0.0599±0.0135 | 0.0139±0.0109 | 0.0460±0.0184 |
| 18 | | DTU | | 0.056 | | 0.6662±0.0143 | 0.1130±0.0051 | 0.0204±0.0047 | 0.0927±0.0083 |
| 19 | | DTU | | 1 | | 0.6334±0.0443 | 0.0944±0.0200 | 0.0120±0.0088 | 0.0824±0.0209 |
| 20 | Leave-one-envelope-out | NJU cGrid | 0.333 | 0.185 | | 0.2792±0.0213 | 0.0424±0.0053 | 0.0557±0.0047 | -0.0133±0.0085 |
| 21 | | NJUcGrid | | 1 | | 0.3442±0.0677 | 0.0331±0.0200 | 0.0255±0.0126 | 0.0076±0.0190 |
| 22 | Leave-one-paired-envelope-out | KUL | 0.500 | 0 | $\mathcal{L}_{\Delta PCC}$ | 0.6688±0.0141 | 0.0564±0.0031 | -0.0019±0.0041 | 0.0583±0.0036 |
| 23 | | KUL | | 0.600 | | 0.6399±0.0411 | 0.0424±0.0122 | -0.0038±0.0049 | 0.0462±0.0102 |
| 24 | | KUL | | 1 | | 0.6599±0.0749 | 0.0346±0.0170 | -0.0141±0.0126 | 0.0487±0.0266 |
| 25 | | DTU | | 0.056 | | 0.5719±0.0300 | 0.0517±0.0150 | 0.0038±0.0044 | 0.0479±0.0160 |
| 26 | | DTU | | 1 | | 0.5784±0.0196 | 0.0481±0.0125 | 0.0021±0.0081 | 0.0460±0.0149 |
| 27 | Leave-one-envelope-out | NJU cGrid | 0.333 | 0.185 | | 0.2391±0.0233 | -0.0314±0.008 | -0.0021±0.0077 | -0.0294±0.0081 |
| 28 | | NJU cGrid | | 1 | | 0.4451±0.0781 | 0.0023±0.0223 | -0.0268±0.0193 | 0.0292±0.0165 |

the unattended stimuli that co-occur with the left-out attended stimulus, causing it to predict those unattended envelopes more strongly. This is a known failure mode for models evaluated on entirely out-of-distribution stimulus identities, and further motivates careful dataset design.

In summary, LOPEO effectively prevents overestimation on unbalanced datasets, though it imposes additional constraints that may reduce usable training data in small datasets and is available only for datasets with constant stimuli pairs.

## V. Conclusion

This paper demonstrates that stimulus reconstruction-based DNN AAD decoders produce systematically overestimated decoding accuracy on unbalanced EEG-AAD datasets, as validated across three public datasets. The proposed LOPEO cross-validation strategy addresses this by ensuring that attended–unattended stimulus pairs are fully withheld from the training set. LOPEO imposes a stricter constraint than LOTO and is therefore more challenging, but provides more reliable evaluation of decoder performance on unbalanced data.

Based on these findings, we offer the following recommendations for future EEG-AAD dataset design. First, a counterbalanced or Latin square stimulus assignment is preferred: each audio segment should serve as the attended stream exactly as often as it serves as the unattended stream, across all subjects. Furthermore, evaluating the balance index of a given experimental protocol in advance can effectively guide researchers in mitigating residual imbalance. We also recommend reporting the balance index for any newly proposed dataset.

It should be noted that these findings are specific to stimulus reconstruction-based AAD. Although overestimation arising from EEG long-range temporal correlations has also been identified [11], [12], whether dataset-imbalance-induced biases exist in spatial AAD or linear decoder frameworks remains an important open question for future work.